\documentclass[%
 reprint,
superscriptaddress,
footinbib,
 amsmath,amssymb,
 prl
]{revtex4-2}
 \usepackage{color} 
\usepackage{graphicx}
\usepackage{dcolumn}
\usepackage{bm}
\usepackage{subfigure}
\usepackage{multirow}
\usepackage{hyperref}
\usepackage[mathlines]{lineno}

\begin{document}

\preprint{APS/123-QED}

\title{Observation of the Electromagnetic Dalitz Transition \boldmath{$h_c \rightarrow e^+e^-\eta_c$}}

\author{M.~Ablikim$^{1}$, M.~N.~Achasov$^{10,c}$, P.~Adlarson$^{67}$, S. ~Ahmed$^{15}$, M.~Albrecht$^{4}$, R.~Aliberti$^{28}$, A.~Amoroso$^{66A,66C}$, M.~R.~An$^{32}$, Q.~An$^{63,49}$, X.~H.~Bai$^{57}$
% ...
\\
\vspace{0.2cm}
(BESIII Collaboration)\\
\vspace{0.2cm} {\it
$^{1}$ Institute of High Energy Physics, Beijing 100049, People's Republic of China\\
% ...
$^{4}$ Bochum Ruhr-University, D-44780 Bochum, Germany\\
% ...
$^{10}$ G.I. Budker Institute of Nuclear Physics SB RAS (BINP), Novosibirsk 630090, Russia\\
% ...
$^{15}$ Helmholtz Institute Mainz, Johann-Joachim-Becher-Weg 45, D-55099 Mainz, Germany\\
% ...
$^{28}$ Johannes Gutenberg University of Mainz, Johann-Joachim-Becher-Weg 45, D-55099 Mainz, Germany\\
% ...
$^{32}$ Liaoning Normal University, Dalian 116029, People's Republic of China\\
% ...
$^{49}$ State Key Laboratory of Particle Detection and Electronics, Beijing 100049, Hefei 230026, People's Republic of China\\
% ...
$^{57}$ University of Jinan, Jinan 250022, People's Republic of China\\
% ...
$^{63}$ University of Science and Technology of China, Hefei 230026, People's Republic of China\\
% ...
$^{66}$ University of Turin and INFN, (A)University of Turin, I-10125, Turin, Italy; (B)University of Eastern Piedmont, I-15121, Alessandria, Italy; (C)INFN, I-10125, Turin, Italy\\
$^{67}$ Uppsala University, Box 516, SE-75120 Uppsala, Sweden\\
% ...
$^{c}$ Also at the Novosibirsk State University, Novosibirsk, 630090, Russia\\
% ...
}}

\date{\today}
\begin{abstract}
Using $(27.12\pm 0.14)\times10^8$ $\psi(3686)$ decays and data samples of $e^+e^-$ collisions with $\sqrt{s}$ from 4.130 to 4.780~GeV collected with the BESIII detector, we report the first observation of the electromagnetic Dalitz transition $h_c\to e^+e^-\eta_c$ with a statistical significance of $5.4\sigma$. We measure the ratio of the branching fractions $\frac{\mathcal{B}(h_c\rightarrow e^+e^-\eta_c)}{\mathcal{B}(h_c\rightarrow \gamma \eta_c)}$ separately for {the $h_c$ samples}  produced via $\psi(3686)\to\pi^0h_c$ and $e^+e^-\to\pi^+\pi^-h_c$. The average ratio is  determined to be $(0.59\pm0.10(\text{stat.})\pm0.04(\text{syst.}))\%$, where the uncertainty includes both statistical and systematic components.
\end{abstract}
\maketitle

The charmonium system, composed of a charm quark bound to an anticharm quark ($c\bar{c}$), has played an important role in our understanding of the fundamental theory of the strong interactions between quarks and gluons, Quantum Chromodynamics (QCD). All charmonium states below the open-charm $D\bar{D}$ threshold have been observed experimentally and are well described by potential models~\cite{charmunion1}. However, our knowledge of the P-wave singlet charmonium state, the $h_c(^1P_1)$, is sparse compared to other charmonium resonances. Its best-measured decay mode is the radiative transition $h_c\rightarrow\gamma\eta_c$~\cite{PDG}, while the sum of all other known $h_c$ decay branching fractions is less than $3\%$. Therefore, there is still much to be learned about the decays of this state. 

Searching for new $h_c$ decay modes is important to constrain theoretical models in the charmonium region~\cite{HC1}.
In particular, electromagnetic (EM) Dalitz decays, such as $h_c\to e^+e^-\eta_c$ in which a virtual photon internally converts into an $e^+e^-$ pair, play an important role in revealing the structure of hadrons and their interactions with photons~\cite{APLL}.  

Due to the absence of allowed $E1$ or $M1$ radiative transitions from the $\psi(3686)$, the production of the $h_c$ proceeds through the isospin-violating decay $\psi(3686)\to\pi^0 h_c$, and is therefore highly suppressed.  However, in $e^+e^-$ machines the $h_c$ can also be produced in the process $e^+e^-\to\pi^+\pi^- h_c$, which is found to have a comparable production rate with that of $\psi(3686)\to\pi^0 h_c$ ~\cite{XYZStudy}.  In this case, the large data samples of $e^+e^-$ annihilations collected with the BESIII detector offer an excellent opportunity to explore the EM Dalitz decays of the $h_c$ using a combination of $h_c$ samples produced in both the processes $\psi(3686)\to\pi^0h_c$ and $e^+e^-\to\pi^+\pi^-h_c$.

In this Letter, we report the first observation of the charmonium EM Dalitz decay $h_c\to e^+e^-\eta_c$ via the processes $\psi(3686)\to\pi^0h_c$ (Mode I) and $e^+e^-\to\pi^+\pi^-h_c$ (Mode II). For Mode I, we use a sample of $(27.12\pm 0.14)\times 10^8$ $\psi(3686)$ events~\cite{psi2Sdata}. For Mode II, we use $e^+e^-$ collision data samples taken at center-of-mass energies {$4.130$, $4.160$, $4.210$, $4.230$, $4.237$, $4.246$, $4.260$, $4.270$, $4.280$, $4.290$, $4.315$, $4.340$, $4.360$, $4.380$, $4.400$, $4.420$, $4.440$, $4.740$ and $4.780$} GeV (called ``XYZ data'' hereafter), corresponding to a total integrated luminosity of $10507.44~\text{pb}^{-1}$~\cite{XYZdata1,XYZdata2,XYZdata3,XYZdata4}.  The measurement of the ratio $\frac{\mathcal{B}(h_c\rightarrow e^+e^-\eta_c)}{\mathcal{B}(h_c\rightarrow \gamma\eta_c)}$ is also presented for the first time, where $\eta_c$ is undetected to improve efficiency. The advantage of comparing these two $h_c$ decay channels is that parts of the systematic uncertainties due to tracking, particle identification (PID), the branching fraction of $\psi(3686) \to\pi^0 h_c$, the cross section of $e^+e^-\to\pi^+\pi^- h_c$, and the number of  $\psi(3686)$ events cancel in the ratio.

The design and performance of the BESIII detector is described in detail in Refs.~\cite{BESIIIdetector,BESIIwhite}. Simulated samples including the inclusive and exclusive ones produced with the {\sc geant}4-based~\cite{Gent4} Monte Carlo (MC) package, which includes the geometric description of the BESIII detector and the detector response, are used to determine the detection efficiency and to estimate the backgrounds. The productions of the $h_c$ resonance from $\psi(3686)\to\pi^0 h_c$ and $e^+e^-\to\pi^+\pi^-$ are simulated by HELAMP and PHSP model respectively, both of them from the MC event generator {\sc evtgen}~\cite{EVGEN1,EVGEN2}. The $E1$ transition $h_c\rightarrow\gamma\eta_c$ is modeled with {\sc evtgen}~\cite{EVGEN1,EVGEN2}, with an angular distribution of $1+\text{cos}^2\theta$ in the $h_c$ rest frame, while the EM Dalitz decay $h_c\rightarrow e^+e^-\eta_c$ is simulated by a new generator as described in the Supplementary Material. The known decay modes of the $\eta_c$ resonance are generated by {\sc evtgen}~\cite{EVGEN1,EVGEN2} with branching fractions set to the world average values~\cite{PDG}, and by {\sc lundcharm}~\cite{lundcharm} for the remaining unknown decays.

For both Mode I and Mode II, candidate charged tracks and EM showers are required to satisfy the following common selection criteria. (i) Charged tracks are reconstructed using the tracking information from the main drift chamber (MDC). The distance of the closest approach of every charged track to the $e^+e^-$ interaction point (IP) is required to be within $\pm10$ cm along the beam direction and within 1 cm in the plane perpendicular to the beam direction. The polar angle $\theta$ between the direction of a charged track and that of the beam must satisfy $|\text{cos}\theta|<0.93$ for an effective measurement in the active volume of the MDC. The combined information of the specific ionization energy loss ($\text{d}E/\text{d}x$) in the MDC and the time-of-flight (TOF) in the TOF system is used to calculate PID confidence levels (CLs) for the electron, pion and kaon hypotheses. The particle type with the highest PID CL is assigned to each track.  (ii) EM showers are reconstructed from clusters of deposited energy in the electromagnetic calorimeter (EMC). The shower energies of photon candidates in the EMC must be greater than \mbox{25 MeV} in the barrel region ($|\text{cos}\theta|<0.80$) or greater than 50 MeV in the endcap regions ($0.86 <|\text{cos}\theta|< 0.92$). Showers located in the transition regions between the barrel and the endcap regions are excluded. To avoid showers caused by charged particles, a photon candidate has to be separated by at least $10^\circ$ from any charged track. In order to suppress electronic noise and energy depositions that are unrelated to the event, the EMC time $t$ of the photon candidates must be in coincidence with collision events within the range $0\le t\le 700$ ns.

For \mbox{Mode I},  pairs of photons are accepted as $\pi^0$ candidates in $\psi(3686)\to \pi^0 h_c$ if their reconstructed invariant mass  satisfies $M_{\gamma\gamma} \in [120,145]~\text{MeV}/c^2$. To improve the signal-to-noise ratio, photons related to $\pi^0$ candidates have to be detected in the barrel EMC region with an energy greater than 40 MeV. To reject background, a one-constraint (1C) kinematic fit is performed to the photon pairs, constraining their invariant mass to the nominal $\pi^0$ mass $M_{\pi^0}$~\cite{PDG}, and the best $\pi^0$ candidate is chosen based on the smallest $\chi^{2}_{\rm 1C}$ value.  Background events from $\psi(3686)\rightarrow\pi^+\pi^-J/\psi$ and $\psi(3686)\rightarrow\gamma\gamma J/\psi$ are suppressed by requiring the recoil mass of each $\pi^+\pi^-$ ($\gamma\gamma$) pair in an event to be outside the range of $M_{J/\psi}\pm7$ MeV$/c^2 $ ($M_{J/\psi}\pm 30$ MeV$/c^2$), where $M_{J/\psi}$ is the nominal $J/\psi$ mass~\cite{PDG}. 

For the decay $h_c\to \gamma\eta_c$, the dominant background for the $E1$ photon is from $\pi^0$ decay. To suppress this background, we combine the $E1$ photon candidate with all other photons in the event, and reject the event if any combination has an invariant mass within 15 MeV/$c^2$ of the nominal $\pi^0$ mass.

For the decay $h_c\to e^+e^-\eta_c$, candidates are required to have at least two charged tracks identified as an electron and a positron. An additional criterion $0.5 < E/p < 1.2$ is applied to the track with higher momentum in the $e^+e^-$ pair to further improve the electron identification, where $E$ and $p$ refer to the energy deposition in the EMC and the momentum measured with the MDC, respectively. 

In order to suppress background from $\pi^0\rightarrow\gamma e^+e^-$ in the $\psi(3686)\rightarrow\pi^0h_c, h_c\rightarrow e^+e^-\eta_c$ decay, the invariant mass of $e^+e^-\gamma^\prime$ is required to be outside the range of $M_{\pi^0}\pm 15$ MeV$/c^2$,
where $\gamma^\prime$ is either photon from the $\pi^0$ candidate.
To remove the background from $h_c\to\gamma \eta_c$, where the photon subsequently converts into an $e^+e^-$ pair in the beam pipe or in the inner wall of the MDC, a photon conversion (PC) finder~\cite{GammaConv} is applied to all $e^+e^-$ pairs. The {PC} point is reconstructed using the two charged trajectories in the $x-y$ plane, which is perpendicular to the beam line. The {PC} length $\delta_{xy}$ is defined as the distance between the PC point and (0,0,0) point in the $x-y$ plane. Photon conversion events accumulate at $\delta_{xy} = 3$ cm and $\delta_{xy} = 6.5$ cm corresponding to the positions of the beam pipe and the inner wall of the MDC. A detailed study~\cite{GammaConv} illustrates that the distributions of $\delta_{xy}$ for data and MC simulations are consistent with each other. By requiring $\delta_{xy}<2$ cm, background from photon conversion is removed.

 For Mode II, candidate events are required to have at least two tracks with opposite charge identified as two pions, while the requirements for the $e^+e^-$ pair from the decay $h_c\to e^+ e^- \eta_c$ and the {requirements for the} $E1$ photon from {the decay} $h_c\to\gamma \eta_c$ are the same as those described above for Mode I.

\begin{figure*}
\subfigure{
\label{psip}
	\includegraphics[width=2.2in]{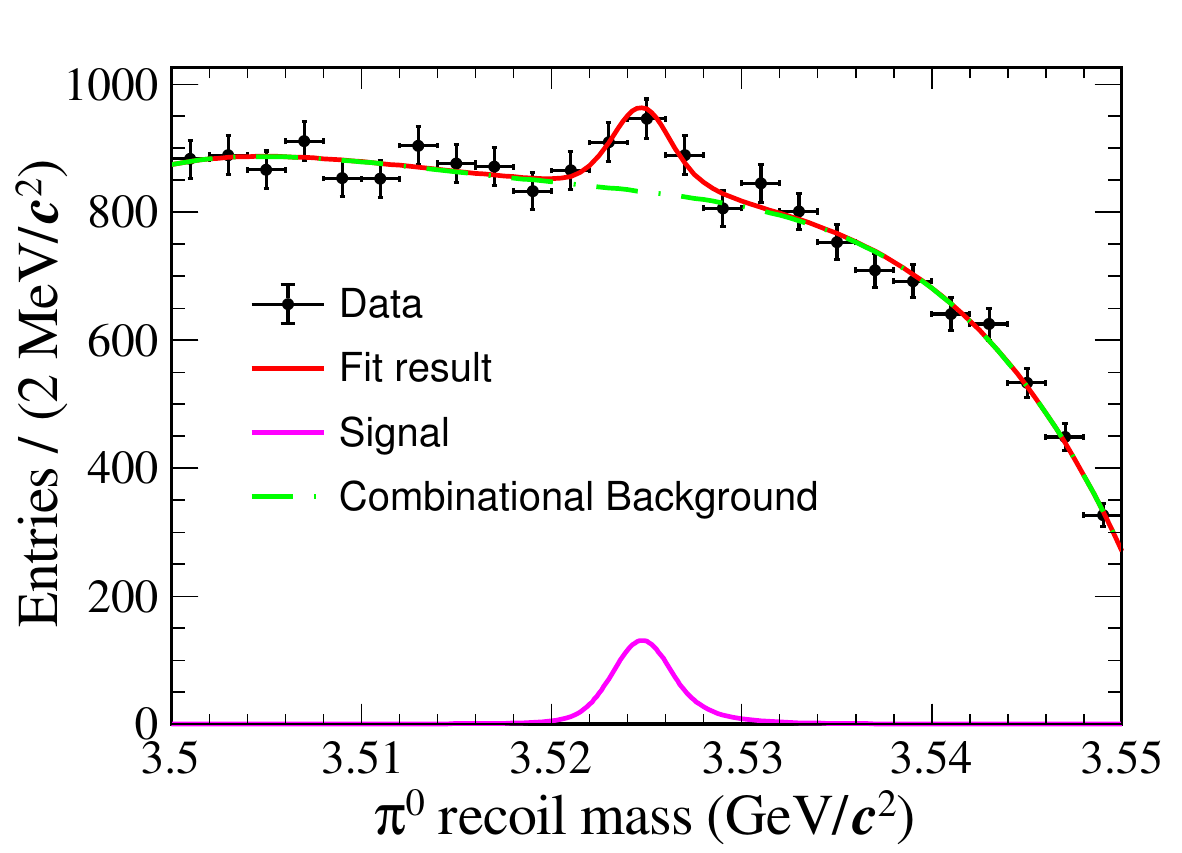}
	\put(-25,95){\scriptsize(a)}
}
\subfigure{
\label{psipGam}
	\includegraphics[width=2.2in]{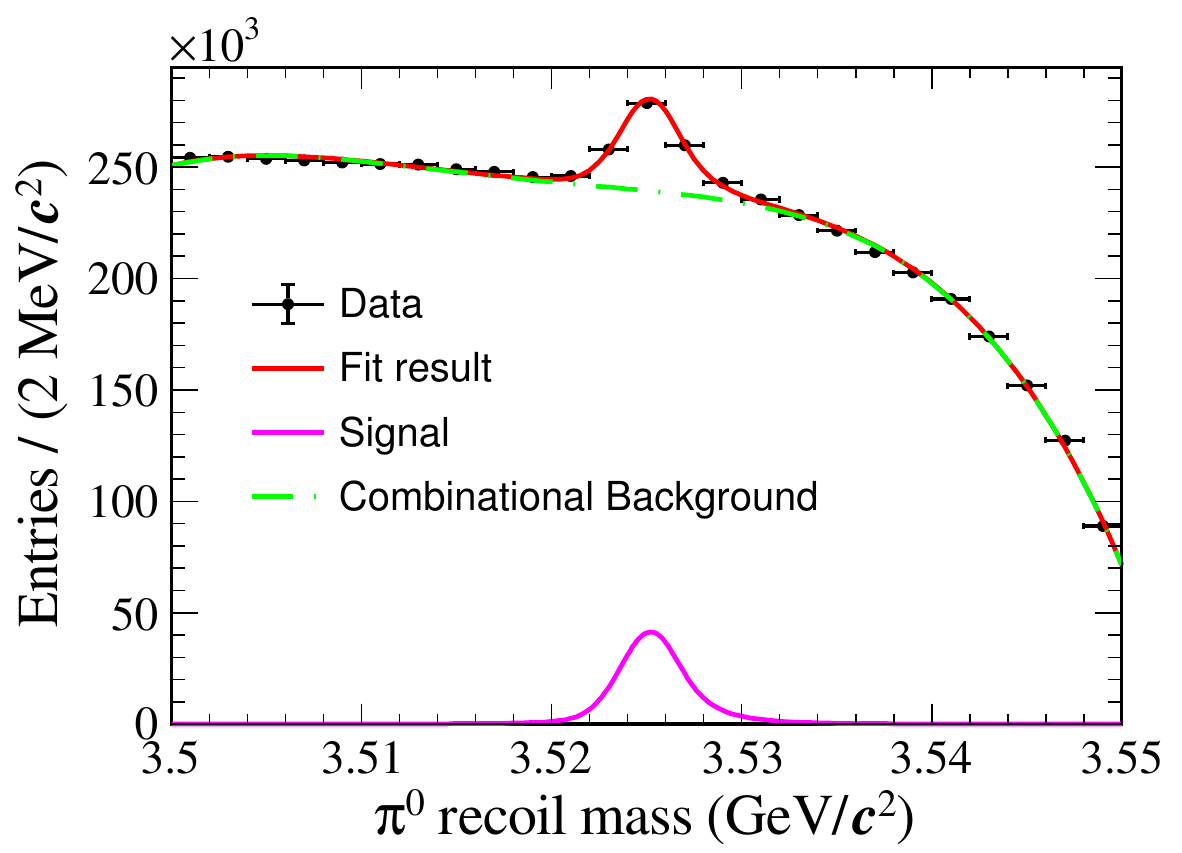}
	\put(-25,95){\scriptsize(b)}
	}
	
\subfigure{
\label{XYZ}
	\includegraphics[width=2.2in]{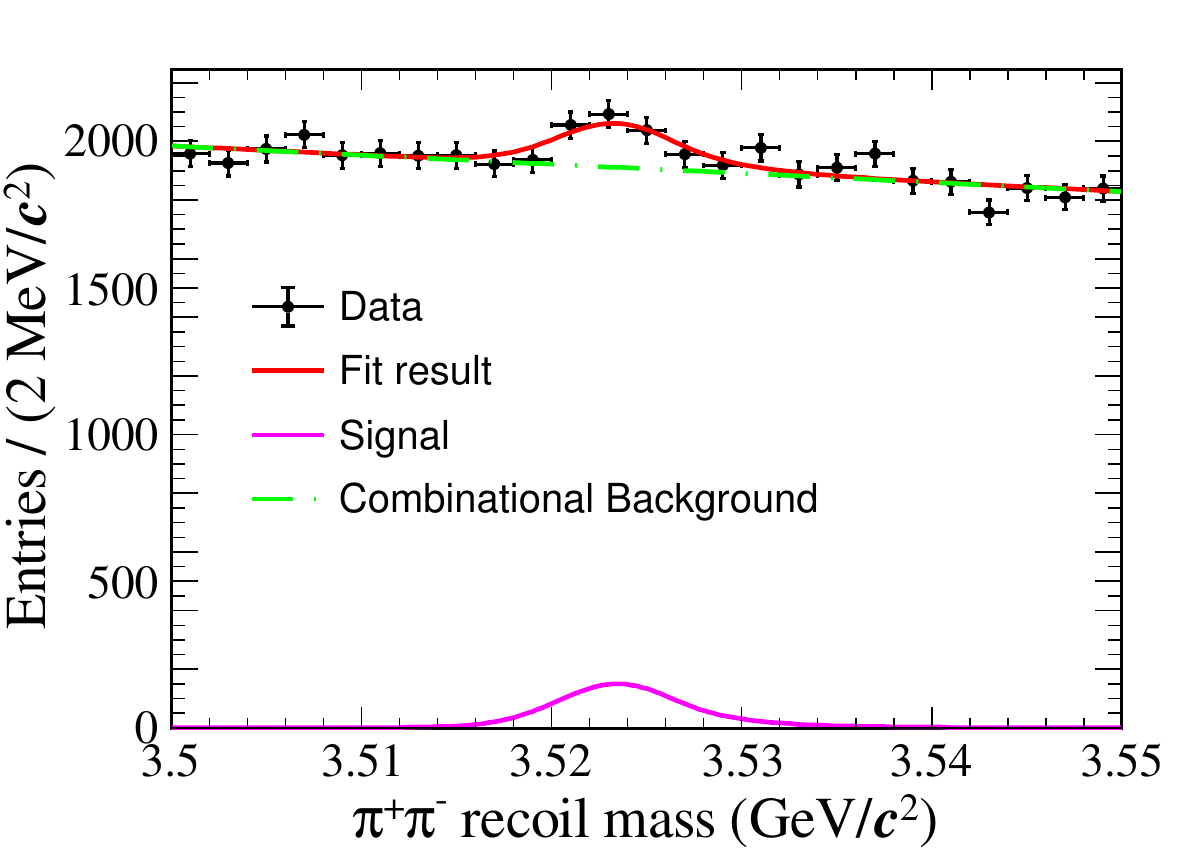}
	\put(-25,95){\scriptsize(c)}
}
\subfigure{
\label{XYZGam}
	\includegraphics[width=2.2in]{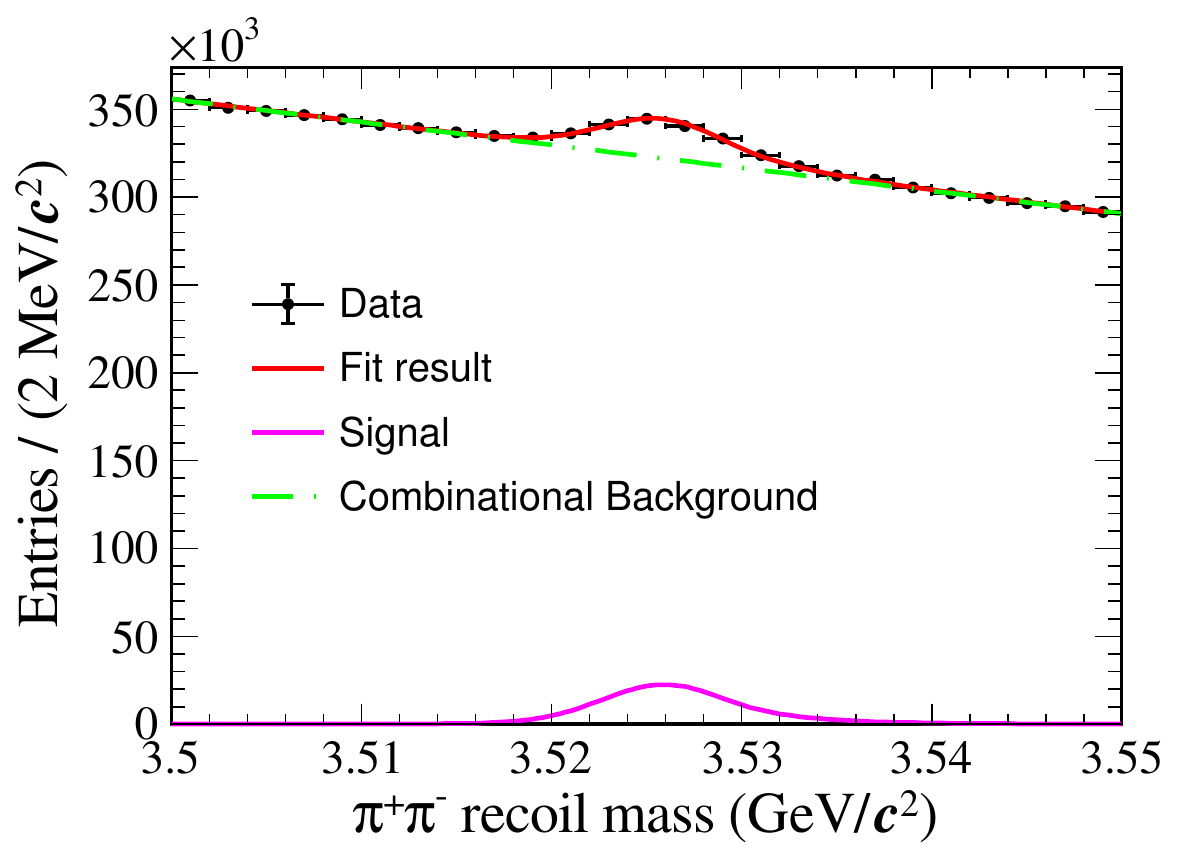}
	\put(-25,95){\scriptsize(d)}
}
\caption{(a,b) The $\pi^0$ recoil mass spectra for $\psi(3686)\rightarrow\pi^0h_c$ {with} (a) $h_c\rightarrow e^+e^-\eta_c$ and (b) $h_c\rightarrow \gamma\eta_c$. (c, d) The $\pi^+\pi^-$ recoil mass spectra for $e^+e^-\rightarrow\pi^+\pi^-h_c$ {with}  (c) $h_c\rightarrow e^+e^-\eta_c$ and  (d) $h_c\rightarrow \gamma\eta_c$  selected from XYZ data. The total fit results are shown as solid red lines, the signals as solid pink lines and the backgrounds as dashed green lines.}\label{fitresults}
\end{figure*}

To select the $\eta_c$ in Mode I, we require that the energy of the $e^+e^-$ pair from $h_c\rightarrow e^+e^-\eta_c$ and the energy of the $\gamma$ from $h_c\rightarrow \gamma\eta_c$ to be in the range $[470, 540]$ MeV. To select the $\eta_c$ in Mode II, we require the masses recoiling against the $\pi^+\pi^- e^+e^-$  and $\pi^+\pi^-\gamma$ systems to be within the $\eta_c$ mass window $[2.92, 3.08]$ GeV$/c^2$.  After the above requirements, extensive examination of the $E_{e^+e^-}$ sideband reveals the absence of any peaking background beneath the $h_c$ signal. The masses recoiling against the $\pi^0$ (\mbox{Mode I}) and $\pi^+\pi^-$ (Mode II) systems are shown in Fig.~\ref{fitresults}. There is a prominent peak for the decays $h_c\to\gamma\eta_c$ and $h_c\to e^+e^-\eta_c$.

To evaluate the statistical significance and the signal yields of the $h_c\to e^+e^-\eta_c$ decay, we perform a simultaneous unbinned maximum likelihood fit to the $\pi^0$ and $\pi^+\pi^-$ recoil mass spectra, which are displayed in Fig.~\ref{psip} and Fig.~\ref{XYZ}. The signal shapes for Mode I and Mode II are parameterized as a common Breit-Wigner function with a fixed width of $0.78$ MeV~\cite{hcmwidth} and a floating resonance mass, convolved with a Crystal ball function to account for the mass resolution effect. While the parameters for the Crystal ball function are fixed to be those obtained from the MC simulations, respectively, for the above two modes.  The background shapes for Mode I and Mode II are described by a fourth-order and a first-order Chebyshev function, respectively.  The fit gives a peak at ($3524.72\pm 0.47)~\text{MeV}/c^2$, which is in good agreement with the world average value of the $h_c$ mass~\cite{PDG}, with a statistical significance of $5.4\sigma$. The statistical significance is determined by the change of the log-likelihood value and the number of degrees of freedom in the fit with and without the $h_c$ signal. To conservatively estimate the significance, the minimum one is ultimately chosen with the effects of both signal and background shapes taken into account. For the reference channel of $h_c\to\gamma\eta_c$, a  simultaneous fit to the $\pi^0$ [Fig.~\ref{psipGam}] and $\pi^+\pi^-$ [Fig.~\ref{XYZGam}] recoil mass spectra is also performed with the same signal model and the background shapes  as those for the $h_c\to e^+e^-\eta_c$ decay.  For the above two cases, the fit results are summarized in Table~\ref{Rcal}.

Using the MC-determined efficiencies and the fitted signal yields, the ratios of the branching fractions, given in Table~\ref{Rcal}, are calculated with

\begin{equation}\label{Rcal}
\mathcal{R}\equiv \frac{\mathcal{B}(h_c\rightarrow e^+e^-\eta_c)}{\mathcal{B}(h_c\rightarrow \gamma\eta_c)} =\frac{N^{\rm obs}_{ e^+e^-\eta_c}}{N^{\rm obs}_{\gamma \eta_c}}\cdot\frac{\epsilon_{\gamma \eta_c}}{\epsilon_{e^+e^-\eta_c}},
\end{equation}
where $N^{\rm obs}_{e^+e^-\eta_c}$ and $N^{\rm obs}_{\gamma\eta_c}$ represent the numbers of the $h_c\to e^+e^-\eta_c$ and $h_c\to \gamma \eta_c$ signal events from the {fits}, and $\epsilon_{e^+e^-\eta_c}$ and $\epsilon_{\gamma \eta_c}$ are the corresponding detection efficiencies. The measured ratios from two Modes are in reasonable agreement within two standard deviations due to the large statistical uncertainties.

\begin{table}
\renewcommand{\arraystretch}{1.5}
\centering
 \caption{The fit results and the measured $\mathcal{R}$ for Mode I and Mode II.}\label{Fit}
 \footnotesize
\noindent
 \begin{tabular}{lccccc}
 \hline\hline
Mode & $N^{\rm obs}_{e^+e^-\eta_c}$&$N^{\rm obs}_{\gamma\eta_c}$&$\epsilon_{e^+e^-\eta_c} $&$\epsilon_{\gamma\eta_c}$ & $\mathcal{R} $(\%)\\
\hline
I & $298\pm76$ & $95564\pm1340$&$5.38\%$&$7.99\%$& $0.46\pm 0.12\pm 0.05$\\
II & $663\pm136$ & $99260\pm1792$ &$24.73\%$&$33.12\%$& $0.89\pm0.18\pm 0.09$\\
 \hline\hline
\end{tabular}
\end{table}

All the systematic uncertainties taken into account are summarized in Table~\ref{totSys}.
\begin{table}
\centering
 \caption{The systematic uncertainties (in \%) in the calculation of $\mathcal{R}$. The items with and without $*$ are common and independent uncertainties, respectively.}\label{totSys}
\noindent
 \begin{tabular}{lcccc}
 \hline\hline
 \multirow{2}*{Source}& \multicolumn{2}{c}{Mode I}&\multicolumn{2}{c}{Mode II}\\
 ~ & $e^+e^-\eta_c$ & $\gamma\eta_c$& $ e^+e^-\eta_c$ & $\gamma\eta_c$\\
 \hline
 Charged tracks* & 2.0 & ---&4.0&2.0 \\
 
 Photon detection* & 2.0 & 3.0 &---&1.0\\
 
 $e^{\pm}$ PID*  & 3.2 & --- &3.2&---\\

 $\gamma$ {conversion} & 1.0&---&1.0&---\\

 $\pi^0$  {reconstruction}*  &  0.7  & 0.7  &---&---\\

Form {factor} & 1.4 &---&1.4&---\\
 
Fitting range & 3.7 &4.7&1.8&0.1\\
 
Background shape &2.4 &2.8 &0.9&6.0\\
 
Signal shape & 5.0 &4.9 &1.8& 5.5\\

 Energy cut & 0.3 &3.9 &---&---\\
 \hline 
Total & \multicolumn{2}{c}{ $11.5$}  &  \multicolumn{2}{c}{$9.6$}\\
 \hline\hline
\end{tabular}
\end{table}
The tracking efficiency has been studied using the control samples of $J/\psi\to e^+e^-(\gamma_{\rm{FSR}})$~\cite{trackingSys} and $J/\psi\to\rho\pi$~\cite{Trackefficicy}, where $\rm FSR$ means final-state radiation. The difference in efficiencies between data and MC simulation for electrons, positrons, and charged pions is estimated to be 1.0\% for each charged track. The total systematic uncertainty is then taken to be 1.0\% times the number of selected charged tracks in each process.

The photon detection efficiency has been studied using a control sample of $J/\psi\rightarrow\rho\pi$~\cite{Trackefficicy}. It is found that the detection efficiency of MC  simulation is in agreement with that of data within  1.0\%, which is taken as the systematic uncertainty for each photon. The total systematic uncertainty from this source is determined by the number of selected good photons in each process.

The uncertainty on electron and positron identification is studied with a control sample of radiative Bhabha scattering events $e^+e^-\rightarrow\gamma e^+e^-$. The average efficiency difference for electron and positron identification between data and MC simulation, weighted according to the polar angle and momentum distributions of the signal MC samples, is determined to be 1.6\% for each of them. Thus, a systematic uncertainty of 3.2\% is assigned.

We perform a study of  $\gamma$ conversion events with a clean control sample of $J/\psi\rightarrow \pi^+\pi^-\pi^0, \pi^0\rightarrow\gamma e^+e^-$. The systematic uncertainty due to  the $\gamma$ conversion veto is estimated to be 1.0\%~\cite{gammaConv2}, which is the relative difference of efficiencies  between data and MC simulation.

The systematic uncertainty due to the $E_{e^+e^-}$ or $E_\gamma$ requirements is estimated by varying the range of $E_{e^+e^-}$ and $E_\gamma$ from $[470,~540]~\text{{MeV}}$ to $[480,~540]~\text{{MeV}}$. The differences  of branching fractions between different energy ranges, 0.3\% and 3.9\%, are taken as the systematic uncertainties from the energy cut of $h_c\to e^+e^-\eta_c$ and $h_c\to \gamma\eta_c$ in $\psi(3686)$ decays, respectively.

The uncertainty due to $\pi^0$ reconstruction is investigated using double-tag $D\bar{D}$ hadronic decay samples of $D^{0} \rightarrow K^{-} \pi^{+}, K^{-} \pi^{+} \pi^{+} \pi^{-}$ {versus} $\bar{D}^{0} \rightarrow K^{+} \pi^{-} \pi^{0}, K_{S}^{0} \pi^{0}$~\cite{pi0eff1,pi0eff2}. The average data-MC difference of the $\pi^0$ reconstruction efficiencies, weighted by the momentum spectra of signal MC events, is 0.7\% per $\pi^0$.

In the generator for the decay $h_c\to e^+e^-\eta_c$, the form factor $V(q^2)=\frac{1}{1-q^2/\Lambda^2}$ is used, where $q^2$ is the square of the invariant mass of the $e^+e^-$ pair, and the pole mass $\Lambda$ is the mass of the vector resonance near the energy scale of the decaying particle according to the Vector Meson Dominance model~\cite{VMD1,VMD2}. To generate the signal MC sample, we assume that the value of $\Lambda$ is $M_{J/\psi}$. In order to obtain the systematic uncertainty from the generator, we change the value of $\Lambda$ to $M_{h_c}$. The difference of detection efficiency gives an uncertainty of 1.4\%.

The fitting range, signal and background descriptions are considered as sources of systematic uncertainty related to the fitting procedure. These uncertainties are determined by varying the fitting ranges, changing the $h_c$ signal shape, and varying the orders of the Chebychev functions in the fit.  The maximum differences of the fit results are assigned as the systematic uncertainties from each item of the fitting procedure as shown in Table~\ref{totSys}.

The systematic uncertainties from different sources and their corresponding contributions are summarized in Table~\ref{totSys}, where the total systematic uncertainties of $\mathcal {R}$ for Mode I and Mode II are calculated by assuming that the common systematic uncertainties between the two $h_c$ decays cancel in the measurement of $\mathcal{R}$.

In summary, based on $(27.12\pm 0.14)\times 10^8$ $\psi(3686)$ events and XYZ data, corresponding to a total integrated luminosity of $10507.44$ $\text{pb}^{-1}$ taken at center-of-mass energies between 4.130 and 4.780 GeV collected with the BESIII detector, we observe the EM Dalitz decay $h_c\rightarrow e^+e^-\eta_c$ with a statistical significance of $5.4\sigma$. We measure the ratio $\mathcal{R}=\frac{\mathcal{B}(h_c\rightarrow e^+e^-\eta_c)}{\mathcal{B}(h_c\rightarrow \gamma\eta_c)}$ separately for samples of $h_c$ produced in the two processes $\psi(3686)\to\pi^0h_c$ and $e^+e^-\to\pi^+\pi^-h_c$, and the results are shown in Table~\ref{Rcal}. Combining the results with the weighted least squares method from $\psi(3686)$ and XYZ data gives $\mathcal{R}=(0.59\pm0.10(\text{stat.})\pm0.04(\text{syst.}))\%$, where correlations of different sources of systematic uncertainties between the two modes are taken into account.  
This result provides new experimental inputs and vertex information on the interaction of charmonium states with the EM field. This is the first measurement of this branching ratio, and is important to guide models in the description of charmonium decays.

The BESIII Collaboration thanks the staff of BEPCII and the IHEP computing center for their strong support. This work is supported in part by National Key R\&D Program of China under Contracts Nos. 2020YFA0406300, 2020YFA0406400, 2023YFA1606000; National Natural Science Foundation of China (NSFC) under Contracts Nos. 11635010, 11735014, 11935015, 11935016, 11935018, 12025502, 12035009, 12035013, 12061131003, 12075250, 12075252, 12192260, 12192261, 12192262, 12192263, 12192264, 12192265, 12221005, 12225509, 12235017, 12361141819; the Chinese Academy of Sciences (CAS) Large-Scale Scientific Facility Program; the CAS Center for Excellence in Particle Physics (CCEPP); Joint Large-Scale Scientific Facility Funds of the NSFC and CAS under Contract No. U1832207; 100 Talents Program of CAS; The Institute of Nuclear and Particle Physics (INPAC) and Shanghai Key Laboratory for Particle Physics and Cosmology; German Research Foundation DFG under Contracts Nos. FOR5327, GRK 2149; Istituto Nazionale di Fisica Nucleare, Italy; Knut and Alice Wallenberg Foundation under Contracts Nos. 2021.0174, 2021.0299; Ministry of Development of Turkey under Contract No. DPT2006K-120470; National Research Foundation of Korea under Contract No. NRF-2022R1A2C1092335; National Science and Technology fund of Mongolia; National Science Research and Innovation Fund (NSRF) via the Program Management Unit for Human Resources \& Institutional Development, Research and Innovation of Thailand under Contracts Nos. B16F640076, B50G670107; Polish National Science Centre under Contract No. 2019/35/O/ST2/02907; Swedish Research Council under Contract No. 2019.04595; The Swedish Foundation for International Cooperation in Research and Higher Education under Contract No. CH2018-7756.

\nocite{*}

\bibliography{arxiv}

\providecommand{\noopsort}[1]{}\providecommand{\singleletter}[1]{#1}%
\begin{thebibliography}{25}%
\makeatletter
\providecommand \@ifxundefined [1]{%
 \@ifx{#1\undefined}
}%
\providecommand \@ifnum [1]{%
 \ifnum #1\expandafter \@firstoftwo
 \else \expandafter \@secondoftwo
 \fi
}%
\providecommand \@ifx [1]{%
 \ifx #1\expandafter \@firstoftwo
 \else \expandafter \@secondoftwo
 \fi
}%
\providecommand \natexlab [1]{#1}%
\providecommand \enquote  [1]{``#1''}%
\providecommand \bibnamefont  [1]{#1}%
\providecommand \bibfnamefont [1]{#1}%
\providecommand \citenamefont [1]{#1}%
\providecommand \href@noop [0]{\@secondoftwo}%
\providecommand \href [0]{\begingroup \@sanitize@url \@href}%
\providecommand \@href[1]{\@@startlink{#1}\@@href}%
\providecommand \@@href[1]{\endgroup#1\@@endlink}%
\providecommand \@sanitize@url [0]{\catcode `\\12\catcode `\$12\catcode
  `\&12\catcode `\#12\catcode `\^12\catcode `\_12\catcode `\%12\relax}%
\providecommand \@@startlink[1]{}%
\providecommand \@@endlink[0]{}%
\providecommand \url  [0]{\begingroup\@sanitize@url \@url }%
\providecommand \@url [1]{\endgroup\@href {#1}{\urlprefix }}%
\providecommand \urlprefix  [0]{URL }%
\providecommand \Eprint [0]{\href }%
\providecommand \doibase [0]{https://doi.org/}%
\providecommand \selectlanguage [0]{\@gobble}%
\providecommand \bibinfo  [0]{\@secondoftwo}%
\providecommand \bibfield  [0]{\@secondoftwo}%
\providecommand \translation [1]{[#1]}%
\providecommand \BibitemOpen [0]{}%
\providecommand \bibitemStop [0]{}%
\providecommand \bibitemNoStop [0]{.\EOS\space}%
\providecommand \EOS [0]{\spacefactor3000\relax}%
\providecommand \BibitemShut  [1]{\csname bibitem#1\endcsname}%
\let\auto@bib@innerbib\@empty
\bibitem [{\citenamefont {Barnes}\ \emph {et~al.}(2005)\citenamefont {Barnes},
  \citenamefont {Godfrey},\ and\ \citenamefont {Swanson}}]{charmunion1}%
  \BibitemOpen
  \bibfield  {author} {\bibinfo {author} {\bibfnamefont {T.}~\bibnamefont
  {Barnes}}, \bibinfo {author} {\bibfnamefont {S.}~\bibnamefont {Godfrey}},\
  and\ \bibinfo {author} {\bibfnamefont {E.~S.}\ \bibnamefont {Swanson}},\
  }\href {https://doi.org/10.1103/PhysRevD.72.054026} {\bibfield  {journal}
  {\bibinfo  {journal} {Phys.\ Rev. D}\ }\textbf {\bibinfo {volume} {72}},\
  \bibinfo {pages} {054026} (\bibinfo {year} {2005})}\BibitemShut {NoStop}%
\bibitem [{\citenamefont {Workman}\ \emph {et~al.}(2022)\citenamefont {Workman}
  \emph {et~al.}}]{PDG}%
  \BibitemOpen
  \bibfield  {author} {\bibinfo {author} {\bibfnamefont {R.~L.}\ \bibnamefont
  {Workman}} \emph {et~al.} (\bibinfo {collaboration} {Particle Data Group}),\
  }\href {https://doi.org/10.1093/ptep/ptac097} {\bibfield  {journal} {\bibinfo
   {journal} {Prog. Theor. Exp. Phys.}\ }\textbf {\bibinfo {volume} {2022}},\
  \bibinfo {pages} {083C01} (\bibinfo {year} {2022})}\BibitemShut {NoStop}%
\bibitem [{\citenamefont {{\relax Barnes, S. {\relax Godfrey}}}\ and\
  \citenamefont {{\relax Swanson}}(2005)}]{HC1}%
  \BibitemOpen
  \bibfield  {author} {\bibinfo {author} {\bibfnamefont {T.}~\bibnamefont
  {{\relax Barnes, S. {\relax Godfrey}}}}\ and\ \bibinfo {author}
  {\bibfnamefont {E.~S.}\ \bibnamefont {{\relax Swanson}}},\ }\href
  {https://doi.org/https://doi.org/10.1103/PhysRevD.72.054026} {\bibfield
  {journal} {\bibinfo  {journal} {Phys.\ Rev. D}\ }\textbf {\bibinfo {volume}
  {72}},\ \bibinfo {pages} {1054026} (\bibinfo {year} {2005})}\BibitemShut
  {NoStop}%
\bibitem [{\citenamefont {Landsberg}(1985)}]{APLL}%
  \BibitemOpen
  \bibfield  {author} {\bibinfo {author} {\bibfnamefont {L.}~\bibnamefont
  {Landsberg}},\ }\href {https://doi.org/10.1016/0370-1573(85)90129-2}
  {\bibfield  {journal} {\bibinfo  {journal} {Phys. Rep.}\ }\textbf {\bibinfo
  {volume} {128}},\ \bibinfo {pages} {301} (\bibinfo {year}
  {1985})}\BibitemShut {NoStop}%
\bibitem [{\citenamefont {Ablikim}\ \emph {et~al.}(2013)\citenamefont {Ablikim}
  \emph {et~al.}}]{XYZStudy}%
  \BibitemOpen
  \bibfield  {author} {\bibinfo {author} {\bibfnamefont {M.}~\bibnamefont
  {Ablikim}} \emph {et~al.} (\bibinfo {collaboration} {BESIII Collaboration}),\
  }\href {https://doi.org/10.1103/PhysRevLett.111.242001} {\bibfield  {journal}
  {\bibinfo  {journal} {Phys.\ Rev.\ Lett.}\ }\textbf {\bibinfo {volume}
  {111}},\ \bibinfo {pages} {242001} (\bibinfo {year} {2013})}\BibitemShut
  {NoStop}%
\bibitem [{\citenamefont {Ablikim}\ \emph {et~al.}()\citenamefont {Ablikim}
  \emph {et~al.}}]{psi2Sdata}%
  \BibitemOpen
  \bibfield  {author} {\bibinfo {author} {\bibfnamefont {M.}~\bibnamefont
  {Ablikim}} \emph {et~al.} (\bibinfo {collaboration} {BESIII Collaboration}),\
  }\href@noop {} {\ }\Eprint {https://arxiv.org/abs/2403.06766}
  {arXiv:2403.06766} \BibitemShut {NoStop}%
\bibitem [{\citenamefont {Ablikim}\ \emph {et~al.}(2015)\citenamefont {Ablikim}
  \emph {et~al.}}]{XYZdata1}%
  \BibitemOpen
  \bibfield  {author} {\bibinfo {author} {\bibfnamefont {M.}~\bibnamefont
  {Ablikim}} \emph {et~al.} (\bibinfo {collaboration} {BESIII Collaboration}),\
  }\href {https://iopscience.iop.org/article/10.1088/1674-1137/39/9/093001}
  {\bibfield  {journal} {\bibinfo  {journal} {Chin.\ Phys.\ C}\ }\textbf
  {\bibinfo {volume} {39}},\ \bibinfo {pages} {093001} (\bibinfo {year}
  {2015})}\BibitemShut {NoStop}%
\bibitem [{\citenamefont {Ablikim}\ \emph {et~al.}(2021)\citenamefont {Ablikim}
  \emph {et~al.}}]{XYZdata2}%
  \BibitemOpen
  \bibfield  {author} {\bibinfo {author} {\bibfnamefont {M.}~\bibnamefont
  {Ablikim}} \emph {et~al.} (\bibinfo {collaboration} {BESIII Collaboration}),\
  }\href {https://iopscience.iop.org/article/10.1088/1674-1137/ac1575}
  {\bibfield  {journal} {\bibinfo  {journal} {Chin. Phys. C}\ }\textbf
  {\bibinfo {volume} {45}},\ \bibinfo {pages} {103001} (\bibinfo {year}
  {2021})}\BibitemShut {NoStop}%
\bibitem [{\citenamefont {Ablikim}\ \emph
  {et~al.}(2022{\natexlab{a}})\citenamefont {Ablikim} \emph
  {et~al.}}]{XYZdata3}%
  \BibitemOpen
  \bibfield  {author} {\bibinfo {author} {\bibfnamefont {M.}~\bibnamefont
  {Ablikim}} \emph {et~al.} (\bibinfo {collaboration} {BESIII Collaboration}),\
  }\href {https://iopscience.iop.org/article/10.1088/1674-1137/ac80b4}
  {\bibfield  {journal} {\bibinfo  {journal} {Chin. Phys. C}\ }\textbf
  {\bibinfo {volume} {46}},\ \bibinfo {pages} {113002} (\bibinfo {year}
  {2022}{\natexlab{a}})}\BibitemShut {NoStop}%
\bibitem [{\citenamefont {Ablikim}\ \emph
  {et~al.}(2022{\natexlab{b}})\citenamefont {Ablikim} \emph
  {et~al.}}]{XYZdata4}%
  \BibitemOpen
  \bibfield  {author} {\bibinfo {author} {\bibfnamefont {M.}~\bibnamefont
  {Ablikim}} \emph {et~al.} (\bibinfo {collaboration} {BESIII Collaboration}),\
  }\href {https://iopscience.iop.org/article/10.1088/1674-1137/ac84cc}
  {\bibfield  {journal} {\bibinfo  {journal} {Chin. Phys. C}\ }\textbf
  {\bibinfo {volume} {46}},\ \bibinfo {pages} {113003} (\bibinfo {year}
  {2022}{\natexlab{b}})}\BibitemShut {NoStop}%
\bibitem [{\citenamefont {Ablikim}\ \emph {et~al.}(2010)\citenamefont {Ablikim}
  \emph {et~al.}}]{BESIIIdetector}%
  \BibitemOpen
  \bibfield  {author} {\bibinfo {author} {\bibfnamefont {M.}~\bibnamefont
  {Ablikim}} \emph {et~al.} (\bibinfo {collaboration} {BESIII Collaboration}),\
  }\href {https://doi.org/10.1016/j.nima.2009.12.050} {\bibfield  {journal}
  {\bibinfo  {journal} {Nucl. Instrum. Methods Phys. Res., Sect. A}\ }\textbf
  {\bibinfo {volume} {614}},\ \bibinfo {pages} {345} (\bibinfo {year}
  {2010})}\BibitemShut {NoStop}%
\bibitem [{\citenamefont {\relax Ablikim~\textit{et al}.}(2020)}]{BESIIwhite}%
  \BibitemOpen
  \bibfield  {author} {\bibinfo {author} {\bibfnamefont {M.}~\bibnamefont
  {\relax Ablikim~\textit{et al}.}} (\bibinfo {collaboration} {BESIII
  Collaboration}),\ }\href {https://doi.org/10.1088/1674-1137/44/4/040001}
  {\bibfield  {journal} {\bibinfo  {journal} {Chin.\ Phys. C}\ }\textbf
  {\bibinfo {volume} {44}},\ \bibinfo {pages} {040001} (\bibinfo {year}
  {2020})}\BibitemShut {NoStop}%
\bibitem [{\citenamefont {Agostinelli}\ \emph {et~al.}(2003)\citenamefont
  {Agostinelli} \emph {et~al.}}]{Gent4}%
  \BibitemOpen
  \bibfield  {author} {\bibinfo {author} {\bibfnamefont {S.}~\bibnamefont
  {Agostinelli}} \emph {et~al.} (\bibinfo {collaboration} {GEANT4
  Collaboration}),\ }\href {https://doi.org/10.1016/S0168-9002(03)01368-8}
  {\bibfield  {journal} {\bibinfo  {journal} {Nucl. Instrum. Methods Phys.
  Res., Sect. A}\ }\textbf {\bibinfo {volume} {506}},\ \bibinfo {pages} {250}
  (\bibinfo {year} {2003})}\BibitemShut {NoStop}%
\bibitem [{\citenamefont {\relax Lange}(2001)}]{EVGEN1}%
  \BibitemOpen
  \bibfield  {author} {\bibinfo {author} {\bibfnamefont {D.~J.}\ \bibnamefont
  {\relax Lange}},\ }\href {https://doi.org/10.1016/S0168-9002(01)00089-4}
  {\bibfield  {journal} {\bibinfo  {journal} {Nucl. Instrum. Methods Phys.
  Res., Sect. A}\ }\textbf {\bibinfo {volume} {462}},\ \bibinfo {pages} {152}
  (\bibinfo {year} {2001})}\BibitemShut {NoStop}%
\bibitem [{\citenamefont {Ping}(2008)}]{EVGEN2}%
  \BibitemOpen
  \bibfield  {author} {\bibinfo {author} {\bibfnamefont {R.~G.}\ \bibnamefont
  {Ping}},\ }\href
  {http://iopscience.iop.org/article/10.1088/1674-1137/32/8/001/meta}
  {\bibfield  {journal} {\bibinfo  {journal} {Chin.\ Phys.\ C}\ }\textbf
  {\bibinfo {volume} {32}},\ \bibinfo {pages} {599} (\bibinfo {year}
  {2008})}\BibitemShut {NoStop}%
\bibitem [{\citenamefont {Yang}\ \emph {et~al.}(2014)\citenamefont {Yang},
  \citenamefont {Ping},\ and\ \citenamefont {Chen}}]{lundcharm}%
  \BibitemOpen
  \bibfield  {author} {\bibinfo {author} {\bibfnamefont {R.~L.}\ \bibnamefont
  {Yang}}, \bibinfo {author} {\bibfnamefont {R.~G.}\ \bibnamefont {Ping}},\
  and\ \bibinfo {author} {\bibfnamefont {H.}~\bibnamefont {Chen}},\ }\href
  {https://cpl.iphy.ac.cn/EN/abstract/article_60019.shtml} {\bibfield
  {journal} {\bibinfo  {journal} {Chin. Phys. Lett.}\ }\textbf {\bibinfo
  {volume} {31}},\ \bibinfo {pages} {061301} (\bibinfo {year}
  {2014})}\BibitemShut {NoStop}%
\bibitem [{\citenamefont {{\relax Xu}}\ and\ \citenamefont {{\relax
  He}}(2012)}]{GammaConv}%
  \BibitemOpen
  \bibfield  {author} {\bibinfo {author} {\bibfnamefont {Z.~R.}\ \bibnamefont
  {{\relax Xu}}}\ and\ \bibinfo {author} {\bibfnamefont {K.~L.}\ \bibnamefont
  {{\relax He}}},\ }\href@noop {} {\bibfield  {journal} {\bibinfo  {journal}
  {Chin.\ Phys.\ C}\ }\textbf {\bibinfo {volume} {36}},\ \bibinfo {pages} {742}
  (\bibinfo {year} {2012})}\BibitemShut {NoStop}%
\bibitem [{\citenamefont {\relax Ablikim~\textit{et al}.}(2022)}]{hcmwidth}%
  \BibitemOpen
  \bibfield  {author} {\bibinfo {author} {\bibfnamefont {M.}~\bibnamefont
  {\relax Ablikim~\textit{et al}.}} (\bibinfo {collaboration} {BESIII
  Collaboration}),\ }\href
  {https://journals.aps.org/prd/abstract/10.1103/PhysRevD.106.072007}
  {\bibfield  {journal} {\bibinfo  {journal} {Phys.\ Rev. D}\ }\textbf
  {\bibinfo {volume} {106}},\ \bibinfo {pages} {072007} (\bibinfo {year}
  {2022})}\BibitemShut {NoStop}%
\bibitem [{\citenamefont {\relax Ablikim~\textit{et al}.}(2017)}]{trackingSys}%
  \BibitemOpen
  \bibfield  {author} {\bibinfo {author} {\bibfnamefont {M.}~\bibnamefont
  {\relax Ablikim~\textit{et al}.}} (\bibinfo {collaboration} {BESIII
  Collaboration}),\ }\href {https://doi.org/10.1103/PhysRevD.96.111101}
  {\bibfield  {journal} {\bibinfo  {journal} {Phys.\ Rev. D}\ }\textbf
  {\bibinfo {volume} {96}},\ \bibinfo {pages} {111101(R)} (\bibinfo {year}
  {2017})}\BibitemShut {NoStop}%
\bibitem [{\citenamefont {\relax Ablikim~\textit{et
  al}.}(2011)}]{Trackefficicy}%
  \BibitemOpen
  \bibfield  {author} {\bibinfo {author} {\bibfnamefont {M.}~\bibnamefont
  {\relax Ablikim~\textit{et al}.}} (\bibinfo {collaboration} {BESIII
  Collaboration}),\ }\href {https://doi.org/10.1103/PhysRevD.83.112005}
  {\bibfield  {journal} {\bibinfo  {journal} {Phys.\ Rev. D}\ }\textbf
  {\bibinfo {volume} {83}},\ \bibinfo {pages} {112005} (\bibinfo {year}
  {2011})}\BibitemShut {NoStop}%
\bibitem [{\citenamefont {\relax Ablikim~\textit{et al}.}(2014)}]{gammaConv2}%
  \BibitemOpen
  \bibfield  {author} {\bibinfo {author} {\bibfnamefont {M.}~\bibnamefont
  {\relax Ablikim~\textit{et al}.}} (\bibinfo {collaboration} {BESIII
  Collaboration}),\ }\href {https://doi.org/10.1103/PhysRevD.89.092008}
  {\bibfield  {journal} {\bibinfo  {journal} {Phys.\ Rev. D}\ }\textbf
  {\bibinfo {volume} {89}},\ \bibinfo {pages} {092008} (\bibinfo {year}
  {2014})}\BibitemShut {NoStop}%
\bibitem [{\citenamefont {\relax Ablikim~\textit{et
  al}.}(2016{\natexlab{a}})}]{pi0eff1}%
  \BibitemOpen
  \bibfield  {author} {\bibinfo {author} {\bibfnamefont {M.}~\bibnamefont
  {\relax Ablikim~\textit{et al}.}} (\bibinfo {collaboration} {BESIII
  Collaboration}),\ }\href {https://doi.org/10.1140/epjc/s10052-016-4198-2}
  {\bibfield  {journal} {\bibinfo  {journal} {Eur. Phys. J. C}\ }\textbf
  {\bibinfo {volume} {76}},\ \bibinfo {pages} {369} (\bibinfo {year}
  {2016}{\natexlab{a}})}\BibitemShut {NoStop}%
\bibitem [{\citenamefont {\relax Ablikim~\textit{et
  al}.}(2016{\natexlab{b}})}]{pi0eff2}%
  \BibitemOpen
  \bibfield  {author} {\bibinfo {author} {\bibfnamefont {M.}~\bibnamefont
  {\relax Ablikim~\textit{et al}.}} (\bibinfo {collaboration} {BESIII
  Collaboration}),\ }\href {https://doi.org/10.1088/1674-1137/40/11/113001}
  {\bibfield  {journal} {\bibinfo  {journal} {Chin.\ Phys. C}\ }\textbf
  {\bibinfo {volume} {40}},\ \bibinfo {pages} {113001} (\bibinfo {year}
  {2016}{\natexlab{b}})}\BibitemShut {NoStop}%
\bibitem [{\citenamefont {\relax Gell-Mann}\ and\ \citenamefont {{\relax
  Zachariasen}}(1961)}]{VMD1}%
  \BibitemOpen
  \bibfield  {author} {\bibinfo {author} {\bibfnamefont {M.}~\bibnamefont
  {\relax Gell-Mann}}\ and\ \bibinfo {author} {\bibfnamefont {F.}~\bibnamefont
  {{\relax Zachariasen}}},\ }\href
  {https://journals.aps.org/pr/abstract/10.1103/PhysRev.124.953} {\bibfield
  {journal} {\bibinfo  {journal} {Phys.\ Rev.}\ }\textbf {\bibinfo {volume}
  {124}},\ \bibinfo {pages} {953} (\bibinfo {year} {1961})}\BibitemShut
  {NoStop}%
\bibitem [{\citenamefont {{\relax Fu, H. B. {\relax Li}, X. S. {Qin}}}\ and\
  \citenamefont {{\relax Yang}}(2012)}]{VMD2}%
  \BibitemOpen
  \bibfield  {author} {\bibinfo {author} {\bibfnamefont {J.~L.}\ \bibnamefont
  {{\relax Fu, H. B. {\relax Li}, X. S. {Qin}}}}\ and\ \bibinfo {author}
  {\bibfnamefont {M.~Z.}\ \bibnamefont {{\relax Yang}}},\ }\href
  {https://doi.org/10.1142/S0217732312502239} {\bibfield  {journal} {\bibinfo
  {journal} {Mod. Phys. Lett. A}\ }\textbf {\bibinfo {volume} {27}},\ \bibinfo
  {pages} {1250223} (\bibinfo {year} {2012})}\BibitemShut {NoStop}%
\end{thebibliography}%

\end{document}


\preprint{APS/123-QED}

\title{Supplementary material for\\
Observation of the Electromagnetic Dalitz Transition \boldmath{$h_c \rightarrow e^+e^-\eta_c$}}

\author{
BESIII Collaboration
}

\date{\today}%

\maketitle

\section{Generator for EM Dalitz decay in charmonium regime}

The process $h_c\rightarrow e^+e^-\eta_c$ is an $A\rightarrow P l^+l^-$ process, where $A$ is an axial-vector meson, $P$ is a pseudoscalar meson, and $l$ is a lepton. The amplitude, $T$, can be written as:
\begin{equation}\label{Amp}
T=\bar{l}\gamma^\mu l\frac{1}{q^2+i\varepsilon}<\eta_c|\bar{c}\gamma_\mu c|h_c>.
\end{equation}
The matrix element $<\eta_c|\bar{c}\gamma_\mu c|h_c>$ is:
\begin{widetext}
 \begin{equation}\label{EQIterm}
<\eta_c|\bar{c}\gamma_\mu c|h_c> = 
i[\varepsilon_{\mu}(m_1+m_2)V_1(q^2)-\frac{\varepsilon\cdot q}{m_1+m_2}(p_1+p_2)_\mu V_2(q^2)-\frac{\varepsilon\cdot q}{q^2}2m_2q_\mu V_3(q^2)]+i\frac{\varepsilon\cdot q}{q^2}2m_2q_\mu V_0(q^2),
\end{equation}
\end{widetext}
where $m_1$ and $m_2$ are the masses of the $h_c$ and $\eta_c$, respectively, and $p_1$ and $p_2$ are their four-momenta; {$\epsilon_{\mu}$ denotes the polarization vector of the $h_c$}; $V_0(q^2)$, $V_1(q^2)$, $V_2(q^2)$ are form factors (single-pole), and $V_3(q^2)=\frac{m_1+m_2}{2m_2}V_1(q^2)-\frac{m_1-m_2}{2m_2}V_2(q^2)$.

Using $p_1^2 = m_1^2$, $p_2^2 = m_2^2$, $p_1=p_2+k_1+k_2$ and $q=p_1-p_2=k_1+k_2$, where $k_1$ and $k_2$ are the four-momenta of the $e^+$ and $e^-$, the final amplitude for $h_c\rightarrow e^+e^-\eta_c$ is written in Eq.~\ref{fianlEQ}, {where $V_1(q^2)^\ast$ and $V_2(q^2)^\ast$ are the complex conjugates of $V_1(q^2)$ and $V_2(q^2)$:}

\begin{widetext}
\begin{eqnarray}\label{fianlEQ}
|T|^2 &=\frac{8}{q^4}\{(m_1+m_2)^2[\frac{(k_1\cdot p_1)(k_2\cdot p_1)}{m_1^2}-k_1\cdot k_2]|V_1(q^2)|^2
+\frac{m_1^2}{(m_1+m_2)^2}[\frac{(p_1\cdot q)^2}{m_1^2}-q^2][\frac{2(k_1\cdot p_1)(k_2\cdot p_1)}{m_1^2}-q^2]|V_2(q^2)|^2\nonumber\\
&+(p_1\cdot q)[k_1\cdot k_2 +m_e^2 - \frac{2(p_1\cdot k_1)(p_1\cdot k_2)}{m_1^2}][V_1(q^2)V_2(q^2)^\ast+V_2(q^2)V_1(q^2)^\ast]\}.
\end{eqnarray} 
\end{widetext}

\nocite{*}